\documentclass[a4paper,12pt]{article}

\usepackage[latin1]{inputenc}
\usepackage[T1]{fontenc}

\usepackage[dvips]{graphicx}
\usepackage{geometry}
\usepackage{verbatim}
\usepackage{amsfonts,amssymb}
\usepackage[normalem]{ulem}
\usepackage{color}
\usepackage{varioref,float,amsmath,amsthm}
\usepackage{mathtools}
\usepackage{multirow}

\begin{document}

\begin{center}
.\\\vspace{7cm}
\noindent\textbf{\LARGE Fractional kinetic modelling of the  adsorption and desorption process from experimental SPR curves}\\ 

\vspace{2cm}
{\small\noindent{\bf Higor V. M. Ferreira$^{1}$, Nelson H. T. Lemes$^{1,*}$, 
Yara L. Coelho$^{2}$, Luciano S. Virtuoso$^{2}$, Ana C. dos Santos Pires$^{3}$  and Luis H. M. da Silva$^{3}$ 
} }\\
\vspace{2cm}
\end{center}

\noindent
\begin{center}
{\small 
$^{1}$ Laboratory of Mathematical-Chemistry, Institute of Chemistry, Universidade Federal de Alfenas (UNIFAL), Alfenas, MG, Brazil  \\
$^{2}$ Colloid Chemistry Group, Institute of Chemistry, Universidade Federal de Alfenas (UNIFAL), Alfenas, MG, Brazil \\
$^{3}$ Applied Molecular Thermodynamics, Department of Food Technology, Universidade Federal de Viçosa (UFV), Viçosa, MG, Brazil\\
$^{4}$ Chemistry Department, Universidade Federal de Viçosa (UFV), Viçosa, MG, Brazil
}
\end{center}
\vspace{\stretch{1}}
\noindent{$^*$nelson.lemes@unifal-mg.edu.br}

\newpage
\subsubsection*{Abstract}

The application of surface plasmon resonance (SPR) has transformed the field of study of interactions between a ligand immobilized on the surface of a sensor chip, designated as $L_S$, and an analyte in solution, referred to as $A$.  
This technique enables the real-time measurement of interactions with high sensitivity. 
The dynamics of
adsorption-desorption process, $A+L_S \rightarrow AL_S$, can be expressed mathematically as a set of coupled integer-order differential equations. However, this approach has limited ability to acoount for temperature distribution, diffusion and transport effects involved in the reaction process. 
The fractional kinetic model provides a methodology for incorporating non-local effects into the problem. In this study, the proposed model was applied to analyze data to the interaction between 
Immobilized Baru Protein (IBP) and Congo Red dye (CR) at concentrations ranging from $7.5$ to 
$97.5$ $\mu M$, at pH $7.4$ and $16^o$ C. 
The variation in the kinetic constants was studied, and it was demonstrated that the integer-order model is unable to adequately represent the experimental data. This work has shown that the fractional-order model is capable of capturing the complexity of the adsorption-desorption process involved in the SPR data.\\

\noindent
{\bf Keywords:} Fractional calculus; adsorption-desorption process; kinetic model; Surface Plasmon Resonance (SPR)  

\clearpage

\section{Introduction}

Surface Plasmon Resonance (SPR) is an optical technique used to measure molecular interactions in real time. The response is quantified in resonance units (RU) and is proportional to the number of analyte molecules bound to the ligand attached to the surface \cite{DOI:10.1021/cr068107d}. Several kinetic models can be employed to describe how reactions occur and the rate of the processes.

The reaction between immobilized ligand ($L_s$) and analyte ($A$) can be assumed to follow pseudo first-order kinetics  when the analyte concentration is approximately constant,  
and the interaction between the ligand and the analyte is
describe by a one-to-one interaction process.{\cite{DOI:10.1006/meth.1994.1013}} 
The determination of the rate constant from the measured data is of great importance, as it allows for the calculation of the thermodynamic interaction parameters. 
 
The dynamics of elementary reactions can be  mathematically expressed as a set of coupled differential equations, which must be integrated simultaneously in order to derive the  concentrations of the 
reactants and products throughout the reaction process.
However, the governing equations in such models are typically analyzed using classical derivatives, 
which offer limited ability to  account for temperature distribution, diffusion, and transport effects involved in the reaction process.

The classical derivative is a local operator, typically taking into account changes in the neighborhood  
of a specific time moment. On the other hand, fractional calculus offers a  
different approach   by  predicting the  behavior of the concentrations of  reactants and products 
over time.
The fractional kinetic model incorporates non-local effects, as the fractional differential operator is defined using integrals that account for changes over 
an interval \cite{ISBN:978-0-444-51832-3}. This property makes fractional derivatives   particularly  suitable for simulating  complex physical phenomena, such as diffusion-reaction processes. 

Recent literature has extended the integer-order kinetic model to include arbitrary orders. For example, reference \cite{DOI:10.1016/j.apm.2016.04.021}, employed 
the fractional differential equation  to model an anomalous luminescence decay process over a long  observation period, 
where deviations from the exponential decay law are expected. 
 
Fractional calculus is a generalization of ordinary calculus, where derivatives and integrals of arbitrary order are defined. The Riemann-Liouville, Caputo and Gr\"unwald-Letnikov definitions are the three most commonly used for fractional differentiation.{\cite{ISBN:9780125588409}}  In physics problems, the Caputo definition is  generally 
preferred, because it provides initial conditions with a clear physical interpretation {\cite{ISBN:978-0-444-51832-3}}.  

Despite 
its growing applications,
many fundamental questions about fractional calculus {remain open, and various calculation results are not directly applicable  to fractional order.}  One of the issues is the dimensionality of the equations, as well as the fact that the fractional derivative of a constant is not always zero {\cite{DOI:10.1007/s40314-024-02680-z,DOI:10.1073/pnas.2322424121,ISBN:9780125588409}}. Furthermore, fractional derivatives still lack a clear geometric and physical  interpretation for arbitrary orders.
Recent studies have shown that fractional calculus is a good alternative tool in various scientific fields, such as the study of the viscoelastic behavior of polymer nanocomposites {\cite{DOI:10.1533/9780857090430.1.184}}, 
modeling the spread of diseases {\cite{DOI:10.1016/j.aej.2024.03.059}}, 
and analyzing the kinetics of drug absorption {\cite{DOI:10.1007/s10928-009-9116-x}}.  
 
 
In this work, a fractional-order differential equation  is presented as an attempt to generalize the usual model 
for analyzing  the SPR experimental data. The kinetic model using the Caputo-type derivative is formulated and solved using operational methods, as well as the Laplace transform and its inverse. The general algebraic solutions are expressed in terms of the Mittag-Leffler function. 

The 
Mittag-Leffler function exhibits behavior that depends on the fractional-order and shows  
different decay rates for short and long times. In fact, the decay is
very fast as $t \rightarrow +0$ and very slow as $t \rightarrow +\infty$ {\cite{DOI:10.3390/e22121359}}. This result 
has been shown to be a convenient way of modeling the experimental SPR data, 
whenever the experimental data show two different behaviors during the 
adsorption and desorption steps. 

In this paper, we present an analysis of the one-to-one kinetic model and its fractional counterpart, discussing their differences and similarities.  Finally, we also provide some examples comparing the classical and fractional models.

{The rest of the paper is organized as follows: 
We present the methodology, starting with the introduction of the 
experimental technique, followed by the presentation of 
 our generalized model and the necessary foundations of 
fractional calculus. The experimental system is described in Section 3. 
In Section 4, the experimental results obtained by the SPR technique are 
reported and analyzed. Finally, the main conclusions are summarized in Section 5.}

\section{Methodology}
\label{sec:met}

\subsection{Surface Plasmon Resonance technique }

SPR is a physical process that occurs when plane-polarized light strikes a thin metal film under conditions of total internal reflection {\cite{ISBN:978-1-53611-857-5,ISBN:978-1-63321-835-2,ISBN:978-1-78262-730-2}}. The SPR technique is a contemporary method for studying  binding equilibrium and kinetics in chemical reactions.  Recently, biosensors have made a significant impact in many fields of 
scientific research, such as antibody-antigen interactions, immunology, virology, and the pharmaceutical industry {\cite{DOI:10.1016/j.trac.2011.01.006,DOI:10.1042/EBC20150010, DOI:10.3390/pr9111996}}. 

There are several configurations of SPR devices that are capable of generating and measuring surface plasmon resonance effect, one of which is the prism-coupled total reflection system. The prism-based SPR system can be applied in different prism arrangements, such as the Kretschmann optical  arrangement {\cite{DOI:10.1515/zna-1968-1247}}. When a beam of light strikes a semicircular prism, the light is deflected towards the interface plane, as it passes from a denser to a less dense medium. By changing the angle of incidence ($\theta$), the angle of reflection of the light changes until a critical angle is reached. At this point, all the light is reflected inside the circular prism. This phenomenon is known as total internal reflection (TIR). 

Although the light is reflected in the direction of the surface, it has an effect in the direction perpendicular to the surface. An electric field is generated in the perpendicular direction, known an evanescent wave, which decreases exponentially with distance from the interface and decays over a distance of about one wavelength of light. The equations that describe how this electric field propagates depend on the refractive index of the medium.%

When the prism is coated with a thin layer of noble metal, the electrons in the conduction band interact with the electric field of the 
evanescent wave. 
Under specific conditions, a resonance effect occurs, leading to maximum energy absorption by the electrons in the metal. 
This forms what is called a surface plasmon.
Surface plasmon is coherent electron oscillations at the interface between any two materials {\cite{DOI:10.1103/PhysRev.106.874}}.
The binding of biomolecules to the sensor surface results in a change in the refractive index. As the electric field (evanescent wave) changes with the refractive index of the medium, the angle of the incident light at which resonance occurs also changes.  

In the SPR technique, the maximum excitation of surface plasmons is detected by monitoring the reflected power as a function of the incident angle. Fortunately, the change in surface refractive index is linear with the number of molecules bound to the sensor surface {\cite{DOI:10.1016/0021-9797(91)90284-F}}. The measured values are converted  to arbitrary units, $RU$ (resonance units), which 
provide information about adsorption-desorption process 
{\cite{ISBN:978-1-53611-857-5,ISBN:978-1-63321-835-2,ISBN:978-1-78262-730-2}}.   

Kinetic parameters of the process can be obtained from non-linear curve fitting analysis of the experimental data. These parameters are of great importance,  as they help to  determine  the thermodynamic interaction values  
between the analyte and the species on the surface {\cite{DOI:10.1110/ps.4330102,DOI:10.1016/j.ijbiomac.2021.06.158}}. 
The selection of the sensor chip is a critical step in the design and execution of a SPR experiment. 
The choice of chip depends on the target to be immobilized on the sensor chip ($L_S$) and the analyte ($A$) flowing over the target to be studied.
 
\subsection{Adsorption-desorption Kinetic Model}
\label{sec:modelx}

SPR has revolutionized the study of the interactions between a ligand immobilized on the surface of a sensor chip, $L_S$, and an analyte in solution, $A$.  
The technique allows real-time measurement of interactions with high sensitivity.

Consider a solution of molecules $A$ in contact with molecules $L_S$ immobilized on a solid surface $S$. The substance adsorbed, $A$, is referred to as the absorbate, while the solid surface $S$ on which the molecules $A$ are immobilized is known as the adsorbent, $A_S$. 
The absorbate can  
bind to the 
solid surface to form an adsorbed complex $AL_S$. 
The attachment of  $A$ to the surface is known as adsorption, while the reverse process is called desorption. The adsorption and desorption processes can be described by the reversible reaction   
$A+L_S= AL_S$, which is mathematically described by the following rate law {\cite{DOI:10.1006/jcis.1998.5771,DOI:10.1006/meth.1994.1013}}  

\begin{equation}
\frac{d{\bf X}}{dt} = F({\bf X};{\bf k})  
\end{equation}

\noindent
where ${\bf X}$ is the time-dependent concentration vector of the species, and ${\bf k}$ represents the 
rate constants vector.  
However, for a proper interpretation of the kinetic process, it is important to consider  mass transport processes, including diffusion and drift, in which adsorbate molecules are introduced into a controlled flow system.  By combining the diffusion and rate equations, the general equation describing a diffusion-reaction process can be written as follows:

\begin{equation}
\frac{\partial {\bf X}}{\partial t} = D\nabla^2 {\bf X}+\kappa \nabla {\bf X} + F({\bf X};{\bf k})                                                                              
\end{equation}

\noindent
in which $D$ is the diffusion coefficient {and $\kappa$ is drift constant.} In an initial approach, the diffusion term has been neglected in the description of the process.

During the association phase, the concentration of the complex $AL_S$ increases as a function of time. The reaction between the immobilized ligand ($L_S$) and the analyte ($A$) can be assumed to follow pseudo-first-order kinetics. Therefore, the rate equation describing how the reaction rate depends on the concentration of the $AL_S$ over time is given by \cite{DOI:10.1006/meth.1994.1013} 

\begin{equation}
d[AL_S]/dt= k_a[A][L_S]_0-(k_a[A]+k_d)[AL_S]
\label{eqkin}
\end{equation}

\noindent
In the complex formation equation, $k_a$ is the association rate constant, which describes the rate of complex formation, while $k_d$ is the dissociation rate constant. The equilibrium constant $K_D=k_a/k_d$ describes the stability of the complex. The concentration of $A$ is controlled by the flow through the pump in the cell, and the mathematical function used to control the flow of $A$ in the cell is given by 

\begin{equation}
[A](t) = [A]_0(H(t-t_{on})-H(t-t_{off}))
\label{eqflow}                                                                       
\end{equation}

\noindent
{in which H is Heaviside function}, $t_{on}$ represent the time when the flow is turned on, and $t_{off}$ represents the time when the flow is turned off. The differential Equation (\ref{eqkin}) is coupled with the flow control function in Equation (\ref{eqflow}).


\subsection{Fractional calculus background}

The history of fractional calculus dates back around 1695, when Leibniz first suggested the idea of half-order derivative in his correspondence with L'Hospital {\cite{ISBN:978-3030205232,DOI:10.1007/BFb0067095}}. 
Since then, many definitions have been proposed for the fractional operator; among them, three are widely used: the Gr\"unwald$-$Letnikov, the Riemann$-$Liouville, and the Caputo definitions. 
In this work, we recall and employ the Caputo approach, which is the most frequent used.
The Caputo fractional derivative
is defined
as following {\cite{DOI:10.1111/j.1365-246X.1967.tb02303.x,ISBN:9780125588409}}
\begin{equation}
{_a^*}D{_t^\alpha} f(t)\coloneqq [J^{n-\alpha }D^nf](t)=\frac{1}{\Gamma(n-\alpha)}\int_a^t\frac{f^{(n)}(u)du}{(t-u)^{\alpha -n+1}}
\label{eqdif}
\end{equation}
where $f^{(n)}(u)=\frac{d^nf(u)}{du^n}$ and $\alpha$ is restricted to real numbers, with $n-1\leq \alpha\leq n$ and $t>a>0$ is
the independent variable, in which $a$ is the lower bound of that variable.
Among the three definitions mentioned, the Caputo definition is the
most commonly used in Physical-Chemistry applications. The main advantage of the Caputo
derivative is that it only requires initial conditions expressed in terms of
integer-order derivatives, which represent well-understood features
of physical situations, thereby making it more applicable 
to real-world problems {\cite{DOI:10.1007/978-3-642-14574-2,ISBN:9780125588409}}.

The fractional operator is clearly non-local, since the fractional derivative depends on the lower boundary of the integral.
This contrasts with  the integer-order derivative, which is a local operator.
An important feature of the Caputo fractional derivative is that it has 
a Mittag-Leffler function provides the solution to the simplest fractional differential equation governing relaxation processes {\cite{DOI:10.1007/978-3-662-43930-2}}. 

The Mittag-Leffler function is defined by a power series that depends on the parameter $\alpha$, 
providing a generalization of the exponential function, to which it reduces for $\alpha=1$.
When $0<\alpha<1$, the Mittag-Leffler function exhibits different rates of decay for small and large times {\cite{DOI:10.21711/26755254/rmu202015}}. 
This characteristic motivates the use of this model to analyze SPR experimental data.

Another
properties of the Caputo fractional derivative are {\cite{DOI:10.1016/j.cnsns.2020.105338,DOI:10.1016/j.jcp.2019.03.008,DOI:10.1016/j.jcp.2014.07.019}}:
a) the Caputo fractional derivative of order zero returns the input function; 
b) the dimension of the fractional derivative of order $\alpha$ of the quantity $m$ with respect to $t$ is given by the ratio of the unit of $m$ to the unit of $t$ raised to the power of $\alpha$;
c) the
Caputo fractional derivative of integer-order $n$ gives the same result as the usual differentiation of order $n$;
d) {the Caputo fractional derivative of a constant function is equal to zero, }
and e) the Caputo fractional derivative is an ill-posed  operator, meaning that small errors in input data may yield large errors in the output result.
For a complete review of the Fractional Calculus see, for example, Podlubny's book{\cite{ISBN:9780125588409}}.

\subsection{Fractional kinetic model}
\label{sec:model}

First, we discuss the kinetics for biosensors. Consider the one-to-one
kinetic model for the binding process. The reaction between immobilized ligand ($A_S$) and analyte ($L$) can be assumed to follow a 
pseudo-first-order kinetics when the analyte concentration is constant and the ligand concentration is given by $[L_S](t)=[L_S]_0-[AL_S](t)$. In this case, the kinetics of the process can be described by Equation (\ref{eqkin}). 

In this paper, in order to analyze the experimental data obtained from the SPR experiment, a fractional-order differential equation is presented as an alternative generalization of the usual model,
{\cite{DOI:10.1137/21M1398549}} 

\begin{equation}
c^{1-\alpha}\hspace{.1cm} _0^*D_t^\alpha [AL_S](t)=k_a[A](t)[L_S]_0-(k_a[A](t)+k_d)[AL_S](t)
\label{kincfracx}
\end{equation}

\noindent
in which $_0^*D_t^\alpha\{.\}$ is  fractional derivative operator of order $\alpha$ in Caputo sense,  
and $c=1/s$. The factor $c$ is required to correct the dimensionality of Equation (\ref{kincfracx}) {\cite{DOI:10.1007/s40314-024-02680-z,DOI:10.1073/pnas.2322424121,ISBN:9780125588409}}. 
A novel generalized kinetic model is developed to provide a comprehensive understanding of the adsorption-desorption mechanism, considering nonlocal effects. 

The  analyte concentration  is variable within the sensor chip. Automated flow control is achieved 
through the use of an Arduino device. 
In the mathematical model, the flow of the analyte in the sensor chip is regulated by the function  
$[A](t)=[A]_0(H(t-t_{on})-H(t-t_{off}))$. 
The function $H(t-\tau)$ is a discontinuous function that takes the value of zero for $t<\tau$ and one for $t\geq\tau$. {\cite{ISBN:978-0123846549}}.  Set $t_{on}=0<t<t_{off}=\tau$, $[B_S]_0=R_{max}$ and $[A]_0=C_0$  which leads Equation (\ref{kincfracx}) to the equation

\begin{equation}
c^{1-\alpha}\hspace{.1cm} _0^*D_t^\alpha [AL_S](t)=a-b[AL_S](t)
\label{eq1p}
\end{equation}

\noindent
in which $a=k_aC_0R_{max}$ and $b=k_aC_0+k_d$. By using the operational method, specifically the Laplace transformation, $\mathcal{L}\{.\}$, 
{\cite{ISBN:978-0123846549,DOI:10.1155/2011/298628,DOI:10.1016/j.aml.2009.05.011}}, 

\begin{equation}
\begin{array}{lll}
\mathcal{L}\{_0^*D_t^\alpha f(t)\}=s^\alpha F(s) - \sum_{k=0}^{n-1} D^kJ^{n-\alpha}f(0) s^{n-1-k}$, $n-1<\alpha<n\\
\mathcal{L}\{E_{\alpha}(-at^\alpha)\}=\frac{s^{\alpha}}{s(s^\alpha+a)}\\
\mathcal{L}\{1-E_{\alpha}(-at^\alpha)\}=\frac{a}{s(s^\alpha+a)}
 \end{array} 
\end{equation}

\noindent
the general solution to the first-stage process, Equation (\ref{eq1p}), is given by

\begin{equation}
[AL_S](t)= (a/b) (1-E_{\alpha}(-bt^\alpha)
\label{eq1}
\end{equation}

\noindent
in which $[AL_S](0)=0$. The equilibrium point of the fractional-order is equal to the corresponding integer-order, $[AL_S]_{eq}= a/b$, thus

\begin{equation}
[AL_S](t)= [AL_S]_{eq}(1-E_{\alpha}(-bt^\alpha))
\end{equation}

\noindent
See that the result for $\alpha=1$ is the same as that found for the integer-order. 

Now consider second stage in which $t>\tau$ and $[A]=0$. In this case, Equation (\ref{kincfracx}) becomes

\begin{equation}
c^{1-\alpha}\hspace{.1cm} _0^*D_\xi^\alpha [AL_S]( \xi)=-k_d[AL_S](\xi)
\label{eq2}
\end{equation}

\noindent
in which $\xi=t-\tau$. 
Again,  we solve the above equation using the Laplace transform technique and applying the rules pertinent to the corresponding fractional derivatives, a general solution to the Equation (\ref{eq2}) is given by

\begin{equation}
[AL_S](t)= [AL_S]_{eq}E_{\alpha,1}(-k_d(t-\tau)^\alpha)
\end{equation}

\noindent
In above equation, if $\alpha=1$, we retrieve the result found for the model with integer-order. 

The interaction kinetics were divided into three distinct phases: a) 
Association stage $(0<t<\tau)$, in which $A$ and $L_S$ bind to form the complex $AL_S$; b) Equilibrium stage, 
when $t\approx \tau$, in which the concentration of the complex remains constant; c) Dissociation stage $(t>\tau)$,  in which  the bonds between $A$ and $L_S$ in complex $AL_S$ break.
The dissociation stage starts when the flow of analyte in the sensor chip is replaced by the flow of a buffer solution. 
Each phase contains information about the interaction between the molecules $A$ and $L_S$ in terms of how fast the association or dissociation occurs and the strength of the overall interaction.
Finally, the solution found for each part can be expressed as

\begin{equation}
[AL_S](t)=
\left \{ \begin{array}{lll}
  0, \hspace{.1cm}t<t_{on}\\
     \gamma\times (1-E_{\alpha}(-bt^\alpha)), \hspace{.1cm}0<t<\tau\\
\hspace{0.01cm} [AB_S]_{eq}E_{\alpha}(-k_d(t-\tau)^\alpha),\hspace{.1cm} t>\tau
 \end{array} 
\right.
\label{eqfrac1}
\end{equation}

\noindent
in which $\gamma=[AL_S]_{eq}/(1-E_{\alpha}(-b\tau^\alpha))$ and $E_{\alpha}$ is Mittag-Leffler function with one parameter $\alpha$ {\cite{DOI:10.1007/978-3-662-43930-2}}. 

The Mittag-Leffler function is defined by the following power series, which converges for all arguments $x\in\mathbb{R}$.

\begin{equation}
E_{\alpha} (x) := \sum_{k=0}^{\infty} \frac{x^k}{\Gamma (\alpha k+1)}
\end{equation}

\noindent
Note that if $\alpha=1$, the Mittag-Leffler function reduces to the exponential function $\exp(x)$. 
Thus, the classical kinetic model is recovered when the fractional order is one. 
The aim of this study is to investigate the potential of fractional derivatives for modeling of SPR experimental data.
The following section presents a comparative analysis of the proposed method with existing approaches. 

The experimental data indicate that the association process begins with accelerated growth, exceeding the rate   predicted by an exponential growth model. Later, 
an exponential growth curve better 
represent of the process. The generalization of the kinetic model to fractional-order was motivated by this behavior. Similarly, during the dissociation stage, 
there is a rapid decline  at a faster rate than that predicted by an exponential decay model,  followed by a slower phase, that is well described by an exponential curve.

The exact solution provide by the fractional model is given by
the Mittag-Leffler function, with one parameter ($\alpha$), representing the order of the fractional derivative operator in the kinetic model. 

This work  offers a new approach  to modeling experimental data from SPR experiment. In this context, a variable-order (VO) approach was used to generalize the kinetic model of the adsorption-desorption process. 
Several definitions of variable-order fractional differential  operator can be found in the literature {\cite{DOI:10.48550/arXiv.2102.09932,DOI:10.1515/fca-2019-0003}}.
The first definition is obtained by replacing
a constant-order $\alpha$ with a variable-order $\alpha(t)$. In this approach, all coefficients for
past samples are obtained for present value of the order, and it is expressed as follows:

\begin{equation}
[^*_aD_t^\alpha(t) f](t)\coloneqq \frac{1}{\Gamma(n-\alpha(t)-1)}\int_a^t\frac{f{(n)(u)}du}{(t-u)^{\alpha(t)-n+1}}
\label{eqdif2}
\end{equation}

\noindent
In variable-order Fractional derivative, the order can vary continuously as a function of time.
Several works  suggest
that many problems can be better described by variable-order 
operators than by constant-order operators {\cite{DOI:10.1098/rspa.2019.0498}}. 

These results motivated the application of the variable-order approach to describe the SPR experiment. We seek to determine the most suitable $\alpha(t)$. 

\begin{equation}
\alpha(t)=\alpha_0 H(t)-(\alpha_0-\alpha_1)H(t-\tau)
\end{equation}

\noindent
where $H$ is the Heaviside step function. The following variable-order differential equation 
provides a better description of the experimental data:

\begin{equation}
c^{1-\alpha(t)}\hspace{.1cm} _0^*D_t^{\alpha(t)} [AL_S](t)=k_a[A](t)[L_S]_0-(k_a[A](t)+k_d)[AL_S](t)
\label{kincfrac2}
\end{equation}

\noindent
in which $_0^*D_t^{\alpha(t)}\{.\}$ is  fractional derivative operator, with variable-order $\alpha(t)$, in Caputo sense, and  
$[A](t)=[A]_0(H(t)-H(t-\tau))$
with $c=1/s$. 

Solving the Equation (\ref{kincfrac2}) using the Laplace transform technique and the rules 
for fractional derivatives, the general solution 
is given by

\begin{equation}
[AL_S](t)=\omega(t)(\gamma\times (1-E_{\alpha_0}(-bt^{\alpha_0})))
+(1-\omega(t))([AB_S]_{eq}E_{\alpha_1}(-k_d(t-\tau)^{\alpha_1})),
\end{equation}

\noindent
in which $\omega(t)=H(t)-H(t-\tau)$ and $\gamma=[AL_S]_{eq}/(1-E_{\alpha_0}(-b(\tau)^\alpha_0))$. 
In this equation, an order $\alpha_0$  was used in the association stage, while a different order $\alpha_1$  was applied in the dissociation stage. 

\section{Experimental data}
\label{sec:exp}

The analysis of biomolecular interactions is crucial in both science and industry. To understand these interactions, experimental SPR data are typically used. This paper focuses specifically on SPR data to extract information about the interactions, particularly the corresponding rate constants.

This study used data on the interaction between Immobilized Baru protein (IBP) and Congo Red (CR) dye at concentrations ranging from $7.5$ to $97.5$ $\mu M$, at pH $7.4$ and $16$°C. 

\section{Results and discussions}

The data obtained from an SPR experiment is typically referred to as a sensorgram,
which represents the system's response proportional to the total complex concentration, 
measured over time for various analyte concentrations.  
The sensorgram for the system described in Section \ref{sec:exp} is shown in Figure \ref{fig7}. 

The sensorgram is divided into three distinct parts. During the association phase, the amount of complex formed ($AL_S$) increases as a function of time until it reaches an equilibrium level. 
While the flow is maintained, the system remains in chemical equilibrium, and the complex concentration 
stays constant. 
When the analyte solution is replaced with a buffer solution, the analyte concentrations decreases to zero, 
initiating the dissociation process. After dissociation, the ligand and the analyte
return to their original state as before binding. It is assumed that the system and experimental conditions  are suitable for the kinetic model validity described in Section \ref{sec:modelx}.

Typically, sensorgrams at varying concentrations of $A$ are fitted to obtain the association ($k_a$) and dissociation ($k_d$) constants. 
The experimental data are traditionally analyzed using a simple model fitting
procedure. In the initial stage of the analysis, a nonlinear fit of the form $R(t) = R_{eq}(1 - e^{-bt})$ is performed for values of $t < \tau$. In this fit, the parameter $b$ is determined for each 
analyte concentration $[A]$, with $R_{eq}$ representing the maximum value of the response variable. 
In the subsequent step, under the assumption that $b = k_a[A] + k_d$, the curve $b \times [A]$ is constructed, allowing for the determination of $k_a$ by the slope of the curve.
To obtain the dissociation constant $k_d$, a non-linear fit is performed for 
$R(t-\tau) = R_{eq} \exp(-kd(t-t))$ in which $t > \tau$, thus  $k_d$ is obtained for each $[A]$.
Subsequently, the average value of $k_d$ is taken. This is the standard method 
to fitting these data, as outlined in reference 
{\cite{DOI:10.1021/acs.jpcb.4c04516,DOI:10.1016/j.ijbiomac.2021.07.087,DOI:10.1016/j.ijbiomac.2021.06.158}}. In this case, the values of $k_a$ and $k_d$ were found to be $1342.3 (M\cdot s)^{-1}$ and 
$0.01020 s^{-1}$, respectively. 
This result is designated as  Case 1 in Table \ref{tab:1}.

This paper proposes an alternative method for obtaining the constants. 
Considering that Equation {(\ref{eqfrac1})} describes the three stages of the process, 
where $a(t)=H(t)-H(t-\tau)$,{\bf  $\gamma =R_{eq}/(1-\exp(-b\tau))$}, $R_{eq}=30.34$ and $\tau=140$. 

\begin{equation}
R(t)=a(t)(\gamma \times (1-\exp(-bt)))+(1-a(t))(R_{eq}\exp(-k_d(t-\tau)))
\end{equation}

\noindent
The kinetic constants were determined by minimizing a nonlinear cost function, as
defined in Equation {(\ref{custo})}.
 
\begin{equation}
E(k_a,k_d)=\sum_i((R(t_i)-R_i^{exp})/R_i^{exp})^2)
\label{custo}
\end{equation}

\noindent
In this case, only the sensorgram corresponding to a concentration of $75 \mu M$ of $A$ was used 
in the adjustment process.
To fit the model to the data, the maximum descent method was employed to minimize the sum of the relative least squares, calculated over the entire $R \times t$ curve obtained at a concentration of $75 \mu M$ . 
The values of $k_a$ and $k_d$ obtained from the previous strategy  
were used as a starting point. The resulting values were $k_a=332.5 (M\cdot s)^{-1}$ and $k_d=0.02335 s^{-1}$, which are referred to as Case 2 in Table {\ref{tab:1}}.

The $E$ values for the other concentrations, using the constants derived for $[A] = 75 \mu M$, are also presented in Table {\ref{tab:1}}. It is noteworthy that the results for cases 1 and 2 differ significantly, with Case 2 generally performing better. Table {\ref{tab:1}} shows that the constants $k_a=332.5$ and $k_d=0.02335$ result in reduced $E$ for almost all $[A]$ values between $7.5$ and $97.5uM$. 
The resulting  $E$ in Case 2 was $42.31$, compared to $183$ in Case 1. 

\begin{center}
\begin{figure}[h]
\centering
\includegraphics[scale=.8]{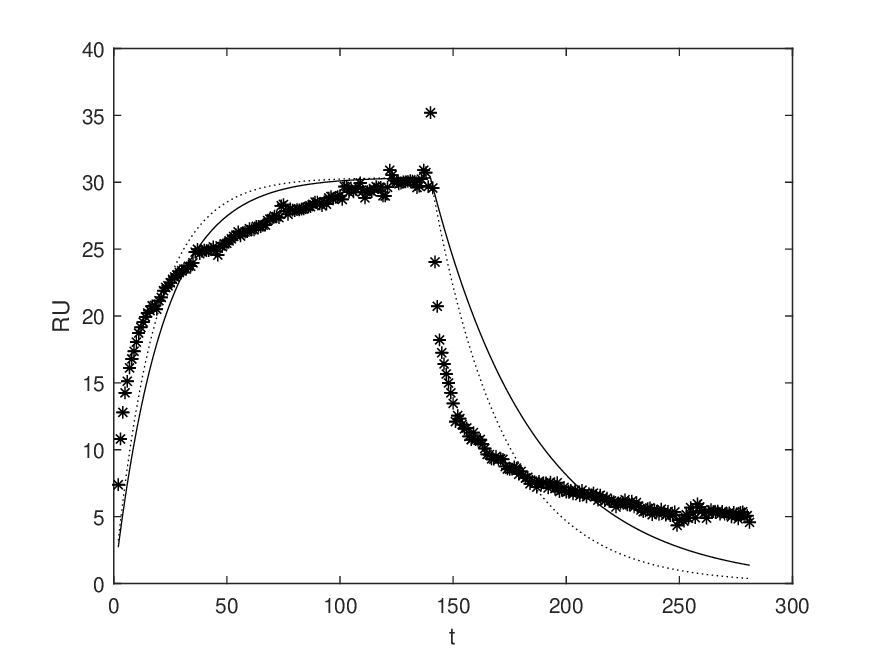}
\caption{Experimental data for an analyte concentration of $75\mu M$ is represented by star symbol. The solid line represents the fit using the strategy labeled Case 1, while the dashed line shows the result obtained from the path labeled Case 2.}
\label{fig3}
\end{figure}
\end{center}

Figure \ref{fig3} illustrates the fit of the integer-order model for both cases, respectively.  
Based on these observations, we conclude that, at least for this experimental data, Case 2 provides a better agreement between the kinetic model and the experimental data than Case 1.

The objective of this study is to determine whether the fractional model offers a more accurate representation of the experimental data. 

\begin{center}
\begin{figure}[h]
\centering
\includegraphics[scale=.8]{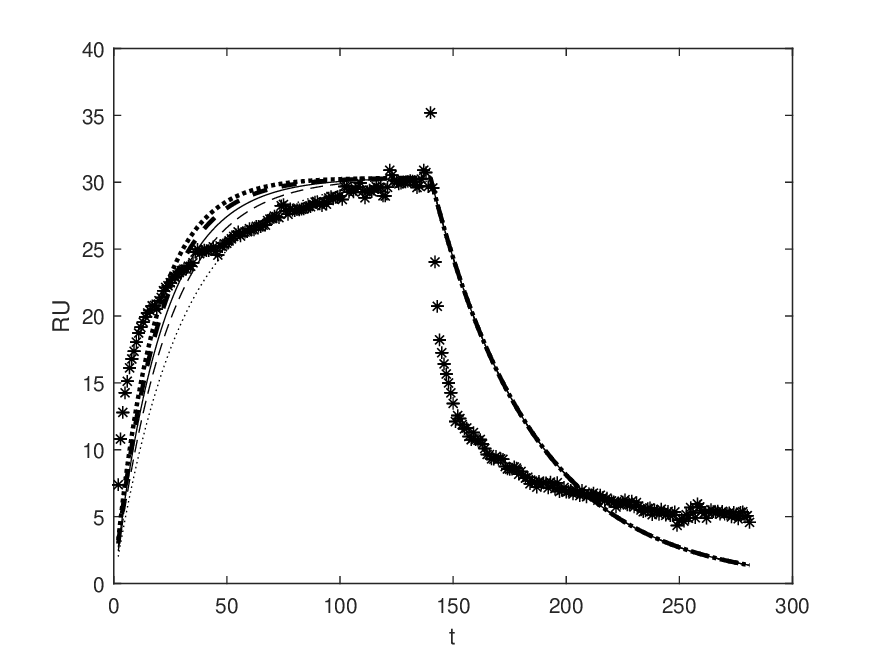}
\caption{Study of the variation of $k_a$. The solid line represents the result obtained by fitting the integer-order model. The thick-dashed line represents the result for a $+20\%$ variation in $k_a$ (with $k_d$ held constant). Similarly, the thin-dashed line represents a $-20\%$ variation in $k_a$. The thick-dotted line represents a $+40\%$ variation in $k_a$, while the thin-dotted line represents a $-40\%$ variation in $k_a$.
}
\label{fig1a}
\end{figure}
\end{center}

Before applying the fractional model, it is essential to examine  the sensitivity of the kinetic constants.
Figure \ref{fig1a} shows a study of the variation of $k_a$.
The solid line represents the result obtained by fitting the integer-order model using reference constants ($k_a=332.5$ and $k_d=0.02335$). The thick-dashed line 
illustrates the effect of a 
$+20\%$ variation in $k_a$ (with $k_d$ held constant). Similarly, the dashed represents line a $-20\%$ variation in $k_a$, while  
the thick-dotted line shows the effect of a $+40\%$ variation in $k_a$, and the dotted line indicates a $-40\%$ variation. 
The simulated changes in $k_a$ primarily affect the curve for  
$t<\tau$, with no significant impact for $t>\tau$. 
This result clearly indicates that adjusting the 
$k_a$  constant alone does not accurately capture the shape of the experimental curve.

\begin{center}
\begin{figure}[h]
\centering
\includegraphics[scale=.8]{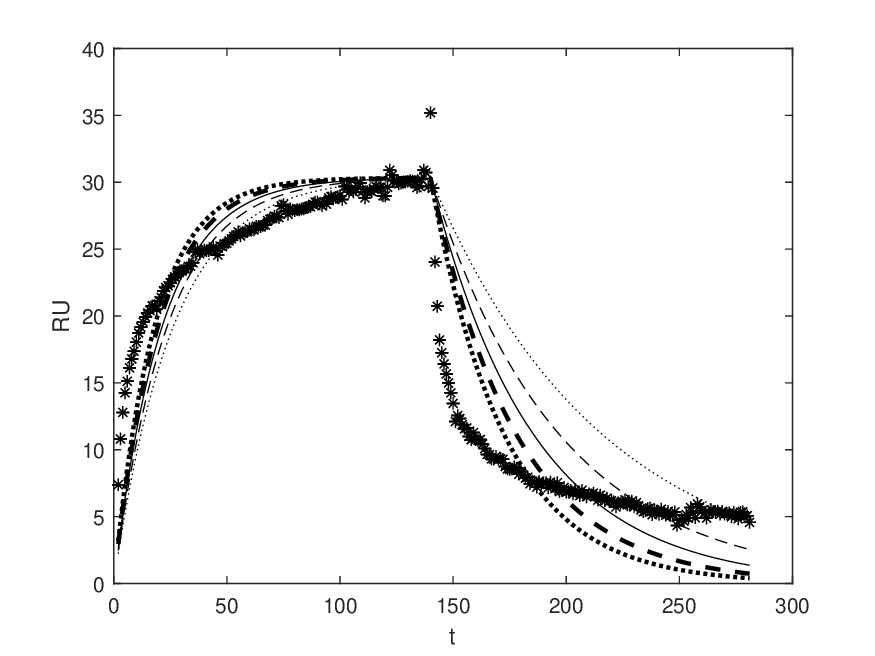}
\caption{The solid line represents the result obtained by fitting the integer-order model. The thick-dashed line represents the result for a $+20\%$ variation in $k_d$, i.e., $k_d=k_d(1+0.2)$ (with $k_a$ held  constant). Similarly, the thin-dashed line represents a $-20\%$ variation in $k_d$. The thick-dotted line represents a $+40\%$ variation in $k_d$, while the thin-dotted line represents a $-40\%$ variation in $k_d$.
}
\label{fig1b}
\end{figure}
\end{center}

Similar study was conducted with $k_d$ constant, as shown in Figure \ref{fig1b}. Again, 
the solid line 
represents the result obtained by fitting the integer-order model using the same reference constants.  
The thick-dashed line corresponds to a $+20\%$ variation in $k_d$, calculated as $k'_d=k_d(1+0.2)$, while keeping $k_a$ constant. 
Similarly, the dashed line represents a $-20\%$ variation in $k_d$, the thick-dotted line indicates a $+40\%$, and the dotted line represents $-40\%$ variation. 
The effects of varying $k_d$ are observed in both regions, $t>\tau$ and $t>\tau$. However, the impact of these changes is more pronounced in the $t>\tau$ region.
As previously noted in the study of $k_a$, modifying the 
$k_d$  constant does not provide adequately represent the empirical curve when using the integer-order model.

Figures \ref{fig1a} and \ref{fig1b} illustrate that the integer-order model fails to accurately represent the experimental curve due to clear deviations from the exponential behavior. 
In the $t<\tau$ region, there is a rapid increase in $R$, which then slows down 
until equilibrium is reached.  In the $t>\tau$ region, 
$R$ decreases rapidly, faster than exponential, followed by a slower decline.   
This behavior suggests 
that a fractional model may be appropriate, as the Mittag-Leffler function is more flexible than the exponential function in accounting for these time-dependent differences.
Therefore, improvements may only be realized by modifying the model.

\begin{center}
\begin{figure}[h]
\centering
\includegraphics[scale=.8]{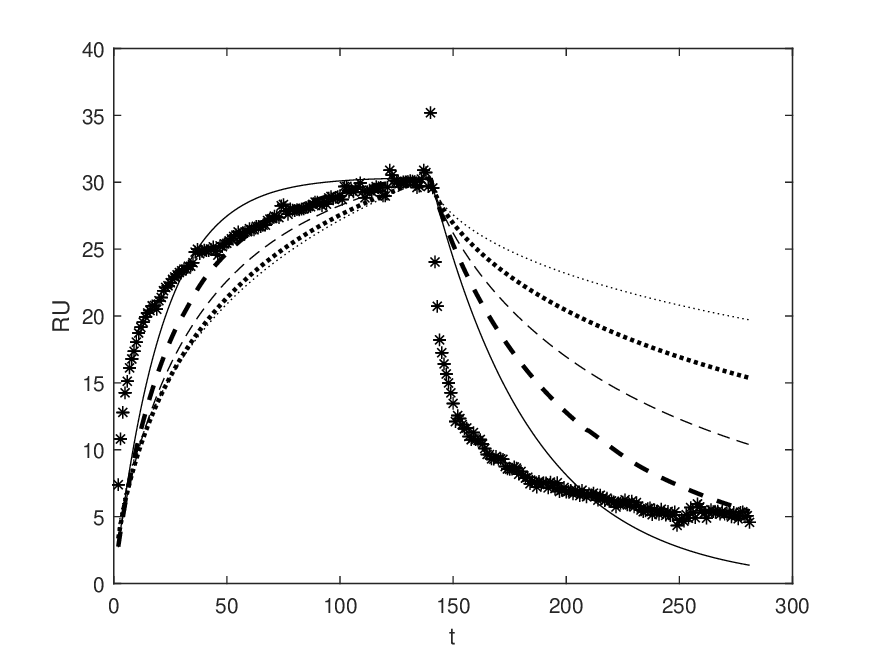}
\caption{
Study of the variation of $\alpha$ (fractional-order), while keeping $k_a$ and $k_d$ fixed.
The solid line represents the result obtained by fitting the integer-order model. The thick-dashed line corresponds to $\alpha=0.9$ (with $k_a$ and $k_d$ held constant). Similarly, the thin-dashed line represents 
$\alpha=0.8$, the thick-dotted line corresponds to $\alpha=0.7$, and the thin-dotted line represents $\alpha=0.6$.
}
\label{fig2}
\end{figure}
\end{center}

An alternative model was employed:

\begin{equation}
R(t)=a(t)(\gamma\times (1-E_{\alpha,1}(-bt^\alpha)))+(1-a(t))(R_{eq}E_{\alpha,1}(-k_d(t-\tau)^\alpha));
\end{equation}

\noindent
where $a(t)$ is defined as before, {\bf $\gamma=R_{eq}/(1-E_{\alpha,1}(-b\tau^\alpha))$}, $R_{eq}=30.34$ and $\tau=140$. 
Figure \ref{fig2} illustrates the effect of varying the fractional-order parameter $\alpha$, with $k_a=332.5 (M\cdot s)^{-1}$ and $k_d=0.02335 s^{-1}$ held constant.
In Figure \ref{fig2}, the solid line represents the fit
using $\alpha=1$. The thick-dashed line corresponds to $\alpha = 0.9$, the dashed line to $\alpha = 0.8$, the thick-dotted line to $\alpha = 0.7$, and the dotted line to $\alpha = 0.6$. 
These curves indicate that varying $\alpha$ while keeping the constants fixed (i.e., a univariate parameter  adjustment) does not improve the fit, as shown in Figure \ref{fig2}. 
Consequently, 
a multivariate adjustment was to implement.
Figure {\ref{fig6}} shows that, for  specific values of  $k_a$ and $k_d$, there is an optimal $\alpha$.
The thick line represents the cost function  as a function of 
$\alpha$, using the $k_a$ and $k_d$ from Case 1. In this case, the 
$\alpha$ value that minimizes $E$ is $0.6$.

\begin{center}
\begin{figure}[h]
\centering
\includegraphics[scale=.8]{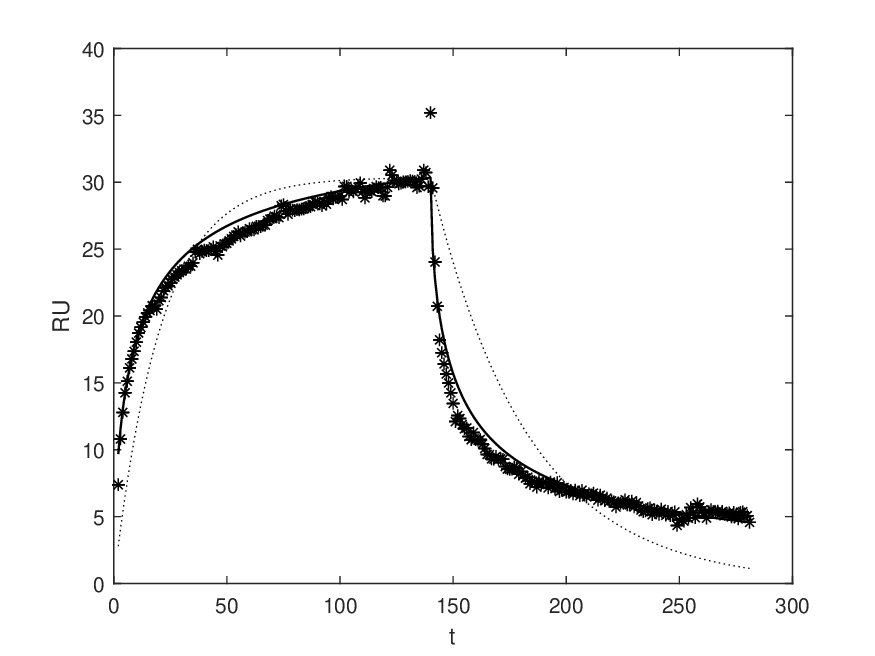}
\caption{Multivariable adjustment of $k_a$, $k_d$, and $\alpha$. Dotted line represents $\alpha=1$ with $E=42.3$. The solid line corresponds 
to $\alpha=0.58$, with $E=1.47$. The star symbols denote the experimental data.}
\label{fig5}
\end{figure}
\end{center}

\begin{center}
\begin{figure}[h]
\centering
\includegraphics[scale=.8]{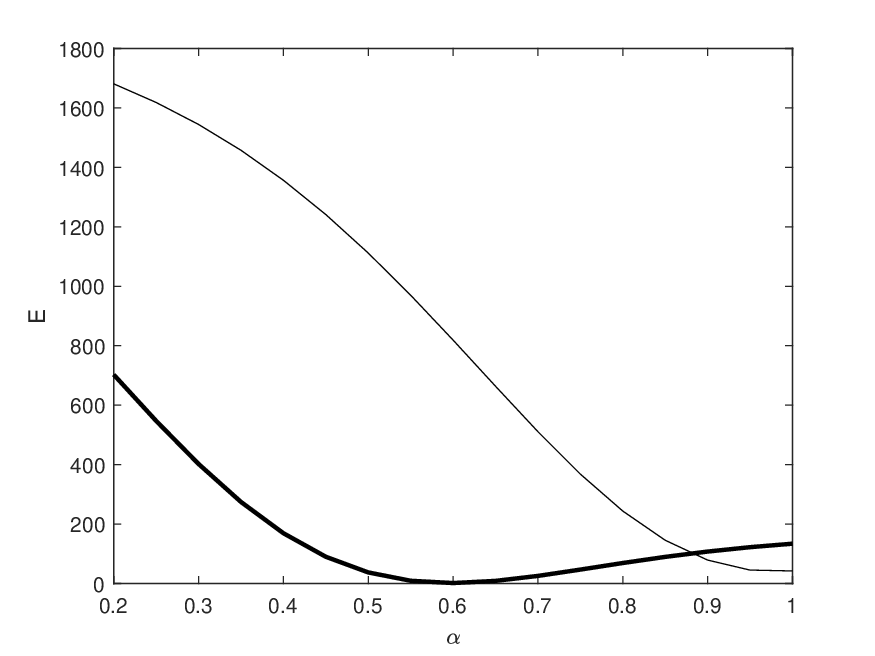}
\caption{
Study of the variation of $\alpha$, keeping $k_a=332.5$ and $k_d=0.02335$ fixed (thin line). Keeping $K_a=332.5$ and $k_d=0.02335 \times 7.9$ fixed (thick line)}
\label{fig6}
\end{figure}
\end{center}

\begin{center}
\begin{figure}[h]
\centering
\includegraphics[scale=.8]{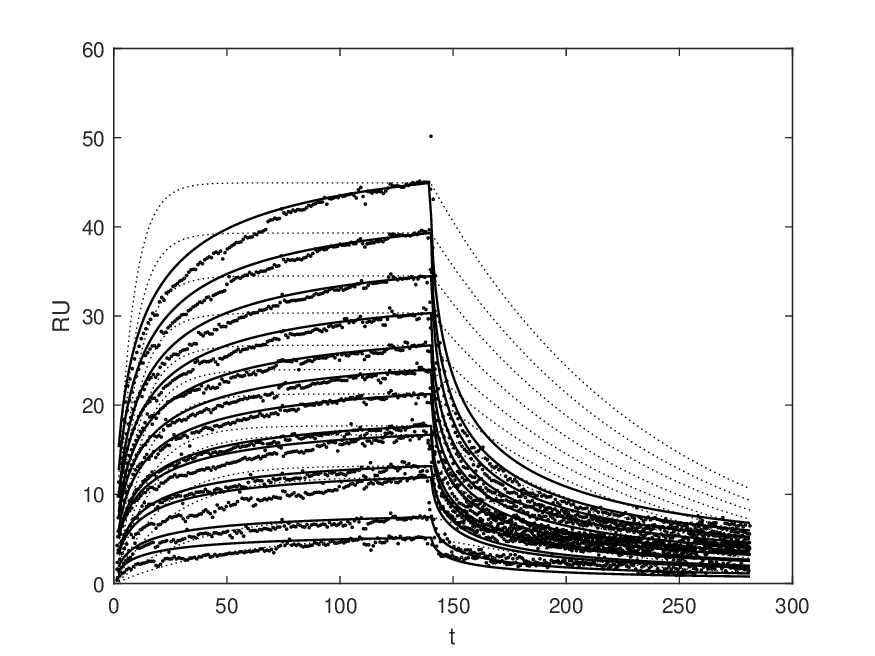}
\caption{The experimental data are compared to the results obtained using the strategy referred to as Case 3 (thick continuous line) and Case 1 (dotted line). }
\label{fig7}
\end{figure}
\end{center}

The next approach involved  a multivariate adjustment by varying $k_a$, $k_d$, 
and $\alpha$, simultaneously, designated as Case 3.  
The same cost function 
$E$ was used.  
Here, the previously determined $k_a$ from the integer-order model was retained,
while $k_d$ was increased by a factor of $7.9$ times, resulting in a new value of $k_d=
0.1844$. Figure \ref{fig5} presents the results. 
The adjustment yielded $\alpha=0.6$ and $E=1.52$, compared to $42.31$ for the integer-order model ($\alpha=1$), reducing the cost function value  by 
a factor of $30$. 

The derivatives of integer-orders can be determined within infinitesimally small neighborhoods of the time point under consideration. This implies that
such equations cannot described non-local effects.
Fractional derivatives of non-integer-orders are employed in kinetic models to describe processes and systems
that exhibit spatial and temporal non-locality, 
the latter often referred to as the ``memory effect". ``Memory" is a property of physical-chemistry processes in which the behavior of a process  at a given specific time $t$ 
depends not only on the current state but on states over a finite or infinite past interval.
This dependence on past states characterizes memory as an inherent property of such processes.
When the influence extends beyond a single location, it is referred to as a non-local effect.

In the context of chemical reactions, interactions between distant molecules, such as dipole-dipole forces or van der Waals forces, can impact reaction processes over large distances. 
The memory effect can be defined as the influence of non-locality on the observed process. 
It has been shown that catalytic surfaces can exhibit memory effects related to the adsorption history of reactive molecules {\cite{DOI:10.1016/j.seppur.2021.120099}}. The state of the surface, including the presence of adsorbed species or morphological changes from previous exposures, 
can affect the efficiency of complex formation in subsequent cycles. Additionally, some  reactions have 
demonstrate memory effects associated with the conformation of the 
ligand immobilized on the surface, which can influence the outcome of subsequent reactions {\cite{DOI:10.1016/j.jcat.2015.11.008}}. In other words, the rate of a reaction at any given moment may depend on 
process history. 

The fractional-order 
provide an 
alternative mean of achieving greater flexibility in the solution while maintaining the simplicity of the 
model. 
The fractional-order that best fit to the experimental curves was  approximately  $0.6$. Figure (\ref{fig7}) presents the sensorgrams for concentrations ranging from $7.5$ to $97.5$ $\mu M$. The experimental data are compared to the results obtained using the Case 3 strategy (thick continuous line). 
For all sensorgrams, optimal results were achieved using the Case 3 strategy, as shown in Table \ref{tab:1}. 


\section{Conclusions}

Over the past few decades, accelerated technological 
advancements have been made in biosensor instruments, such as Surface Plasmon Resonance (SPR) and Quartz Crystal Microbalance (QCM). These techniques enable real-time measurement of biomolecular interactions with high sensitivity.

The adsorption-desorption process observed through the SPR signal is complex and involves several 
concurrent phenomena. In this context, conventional kinetic models
using first-order derivatives to describe the time evolution of the adsorption-desorption process are 
often insufficient for a detailed description. 
In this work, a model employing the Caputo fractional derivative was used to analyze the experimental data. The proposed model was applied to interaction data between Immobilized Baru Protein (IBP) and Congo Red (CR) dye at concentrations ranging from $7.5$ to $97.5$ $\mu M$, at pH 7.4 and 16 $^oC$.  A study of the variation of the kinetic constants showed that the integer-order model fails to adequately describe the experimental data. 

In the filed of chemical reactions, there are examples where interactions between distant molecules, such as dipole-dipole forces or van der Waals forces, can significantly impact reaction processes, even over large distances. Moreover, certain reactions have been observed to exhibit memory effects associated with the conformation of ligands immobilized on surfaces, which can influence subsequent reactions. 
This  means that the rate of a reaction at any given moment may depend on the process's prior history. This study demonstrates that the fractional-order can effectively capture the complexity of such phenomena, offering a robust framework to account for non-local and memory-dependent effects.

The behavior observed in the experimental sensorgrams suggests that the Mittag-Leffler function 
provides greater flexibility than the exponential function in fitting the sensorgram's shape over time.
Using the fractional model, the cost function $E$ was reduced by a factor of 30, with $k_a$ and $k_d$ values comparable to those obtained from the integer-order model. These  
findings highlight the importance of
further investigating the proposed fractional model to better understand the relationship between fractional-order kinetics and non-local phenomena.

\section*{Acknowledgments}

This study was financed in part by the Coordena\c c\~ao de
Aperfei\c coamento de Pessoal de N\'ivel Superior (CAPES) - Finance Code 001, and Funda\c c\~ao de Amparo \`a Pesquisa do Estado de Minas Gerais (FAPEMIG). 



\clearpage
\begin{table}[]
\centering
\caption{The cost function for the three cases}\vspace{.2cm}
\label{tab:my-table}
\begin{tabular}{llll}
\hline\hline
\multicolumn{1}{l|}{\multirow{2}{*}{}} & \multicolumn{3}{l}{E($\alpha$)} \\ \cline{2-4} 
\multicolumn{1}{l|}{$[A],\mu M$}&\multicolumn{1}{l|}{Case 1}&\multicolumn{1}{l|}{Case 2}&\multicolumn{1}{l}{Case 3}\\ \hline\hline
\multicolumn{1}{l|}{7.5}  & \multicolumn{1}{l|}{31.7(1)} & \multicolumn{1}{l|}{42.3(1)} & \multicolumn{1}{l}{19.0(0.5)} \\ 
\multicolumn{1}{l|}{15}  & \multicolumn{1}{l|}{54.8(1)} & \multicolumn{1}{l|}{39.4(1)}  & \multicolumn{1}{l}{22.8(0.55)} \\ 
\multicolumn{1}{l|}{22.5}  & \multicolumn{1}{l|}{18.6(1)} & \multicolumn{1}{l|}{47.1(1)} & \multicolumn{1}{l}{5.75(0.45)} \\ 
\multicolumn{1}{l|}{30}  & \multicolumn{1}{l|}{62.3(1)} & \multicolumn{1}{l|}{39.2(1)}  & \multicolumn{1}{l}{4.25(0.55)} \\ 
\multicolumn{1}{l|}{37.5}  & \multicolumn{1}{l|}{67.0(1)} & \multicolumn{1}{l|}{38.0(1)}  & \multicolumn{1}{l}{2.24(0.55)} \\ 
\multicolumn{1}{l|}{45}  & \multicolumn{1}{l|}{134(1)} & \multicolumn{1}{l|}{40.6(1)}  & \multicolumn{1}{l}{3.37(0.6)} \\ 
\multicolumn{1}{l|}{52.5}  & \multicolumn{1}{l|}{126(1)} & \multicolumn{1}{l|}{40.3(1)}  & \multicolumn{1}{l}{5.33(0.55)} \\ 
\multicolumn{1}{l|}{60}  & \multicolumn{1}{l|}{157(1)} & \multicolumn{1}{l|}{39.2(1)} &  \multicolumn{1}{l}{1.74(0.6)} \\ 
\multicolumn{1}{l|}{67.5}  & \multicolumn{1}{l|}{225(1)} & \multicolumn{1}{l|}{44.8(1)}  & \multicolumn{1}{l}{2.26(0.6)} \\ 
\multicolumn{1}{l|}{75}  & \multicolumn{1}{l|}{183(1)} & \multicolumn{1}{l|}{42.3(1)} &  \multicolumn{1}{l}{1.52(0.6)} \\ 
\multicolumn{1}{l|}{82.5}  & \multicolumn{1}{l|}{214(1)} & \multicolumn{1}{l|}{43.9(1)}  & \multicolumn{1}{l}{3.04(0.6)} \\ 
\multicolumn{1}{l|}{90}  & \multicolumn{1}{l|}{255(1)} & \multicolumn{1}{l|}{45.8(1)} &  \multicolumn{1}{l}{2.56(0.6)} \\ 
\multicolumn{1}{l|}{97.5}  & \multicolumn{1}{l|}{273(1)} & \multicolumn{1}{l|}{45.9(1)} & \multicolumn{1}{l}{3.34(0.6)} \\ 
\hline\hline
\end{tabular}
\label{tab:1}
\end{table}


\clearpage

\end{document}